# Numerical solutions of resistive finite-pressure magnetohydrodynamic equilibria for non-axisymmetric toroidal plasmas

Jian Zhang[a], Ping Zhu[a,b,*] and Chris C. Hegna[b]

[a] State Key Laboratory of Advanced Electromagnetic Technology, International Joint Research Laboratory of Magnetic Confinement Fusion and Plasma Physics, School of Electrical and Electronic Engineering, Huazhong University of Science and Technology, Wuhan, 430074, China

[b] Department of Nuclear Engineering and Engineering Physics, University of Wisconsin-Madison, Madison, WI 53706, United States of America

* Corresponding Author: zhup@hust.edu.cn

Abstract

A hybrid spectral/finite-element code is developed to numerically solve the resistive finite-pressure magnetohydrodynamic equilibria without the necessity of postulating nested magnetic flux surfaces in the non-axisymmetric toroidal systems. The adopted approach integrates a hyperbolic parallel damping equation for pressure updating, along with a dynamic resistive relaxation for magnetic field. To address the non-axisymmetry in toroidal geometry, a pseudo flux mapping is employed to relate the axisymmetric computational domain to the physical domain. On the computational mesh, an isoparametric C1-continuous triangular element is utilized to discretize the poloidal plane, which is complemented with a Fourier decomposition in the toroidal direction. The versatility of the code is demonstrated through its application to several different non-axisymmetric toroidal systems, including the inherently three-dimensional equilibria in stellarators, the helical-core equilibrium states in tokamak plasmas, and the quasi-single-helicity states in a reversed-field pinch.

## 1. Introduction

Toroidal systems, including tokamak, stellarator, and reversed-field pinch (RFP), etc., are believed to be the most promising paths to magnetically controlled nuclear fusion. In earlier studies, the magnetic configuration in tokamaks is regarded as purely axisymmetric and thus two-dimensional, the corresponding equilibrium can be obtained by solving the Grad-Shafranov equation[1,2]. However, many experiments reveal that the actual tokamak equilibria can be three-dimensional (3D) due to the presence of non-axisymmetric instabilities[3,4], or external magnetic perturbations[5]. Besides that, in RFP, another nominally axisymmetric device, the innermost kink-tearing perturbation may saturate and evolve to a helical equilibrium, namely, the quasi-single-helicity (QSH) state[6]. Moreover, the stellarator equilibrium has an intrinsically non-axisymmetric configuration, especially for modern stellarators with a highly optimized boundary[7-9]. These toroidal confinement systems call for equilibrium solvers that are capable of effectively finding the 3D magnetohydrodynamic (MHD) equilibria of resistive plasmas with finite pressure, and the optimized designs of these devices with advanced configurations require reliable and efficient numerical solutions of various distinct equilibrium states in order to evaluate their confinement and stability properties[10-12].

Significant efforts have been made to develop numerical solvers for MHD equilibria in 3D toroidal geometry over the past few decades. Kruskal and Kulsrud[13] first pointed out that the problem of finding a static ideal MHD equilibrium, which satisfies

$$\boldsymbol{J} \times \boldsymbol{B} = \nabla p, \qquad \mu_0 \boldsymbol{J} = \nabla \times \boldsymbol{B}, \qquad \nabla \cdot \boldsymbol{B} = 0,$$

can be converted into a minimization problem for the potential energy of the plasma within nested flux surfaces (NFS). Due to the NFS assumption, the magnetic field can be represented in toroidal flux coordinates and, therefore, obtained by solving the "inverse" equilibrium problem. The VMEC[14] code adopts this approach with an inverse mapping using a combined spectral/finite-difference method and has now become the most widely used 3D toroidal equilibrium solver. Recently, a pseudo-spectral code DESC[15] using global Fourier–Zernike basis functions was developed to solve the equations with the same constraints, which is able to achieve more accurate solutions because of its continuity of higher-order derivatives in the radial direction[16].

However, the NFS assumption leads to the parallel current singularity at rational surfaces with the static ideal MHD equilibrium model[17], and also excludes more general and realistic situations from magnetic islands to stochastic field lines. To deal with this issue, the PIES[18] code splits the magnetic field into two parts, $\boldsymbol{B} = \boldsymbol{B}_0 + \delta \boldsymbol{B}$, where $\boldsymbol{B}_0$ is the main field with NFS and $\delta \boldsymbol{B}$ is a small perturbation that allows the magnetic field to weakly diffuse relative to the unperturbed field. PIES updates the magnetic field from the current density while keeping the pressure unchanged. The perpendicular current density is determined by the force balance equation, and the parallel component is derived from the magnetic differential equation[19]. The SIESTA[20] code is based on the Kulsrud-Kruskal MHD energy minimization principle without the NFS assumption. It expects that the departure of the MHD energy of the final state is small from the initial equilibrium with nested surfaces, thereby the linearized MHD equations can be used to search for the final state with a lower potential energy.

It should be noted that a resistive equilibrium may be far from an ideal one, for instance, the 3D non-sawtoothing stationary states with a central shear-free safety factor in tokamak[21], the resonant QSH states in RFP, and the doublet-like configurations in stellarator[22]. The SPEC[23] code, based on the multi-region relaxed MHD model, allows for much more complicated field behavior. It seeks a weak solution to ideal MHD equilibrium via dividing the total plasma volume into a set of sub-volumes separated by non-relaxed interfaces, and in each sub-volume, the plasma relaxes to the Taylor state[24]. Hence, the ideal MHD equilibrium should be recovered if the number of sub-volumes comes to infinity. The HINT[25,26] code is specially designed to solve for the free-plasma-boundary resistive equilibrium (the location of vacuum-plasma separation is not pre-determined but varies until convergence). It adopts a non-variational relaxation for magnetic fields, and updates pressure by field line tracing, both in a rotational computation box. The vacuum-plasma separation is controlled by priori connection lengths. For simplicity, in the rest of this paper, we use the term "free-" and "fixed-boundary" to refer to the free- and fixed-plasma-boundaries respectively.

Despite the great advances in this field, it is still necessary to develop a new equilibrium solver to meet various requirements for a wide range of non-axisymmetric toroidal plasmas. Some features should be merged and new improvements be considered. The first is, the inclusion of the resistive and finite-pressure effects, and the second is, the accommodation of the desired constraints on fields, such as, the number of toroidal periods, which is essential in stellarator simulations and sometimes used for RFP-QSH equilibrium reconstructions[12]. And also, one may wish to impose the up-down symmetry[27] in some cases. Third, the new code should be capable of solving for equilibria of both fixed- and free-boundaries. Although free boundary equilibrium is more realistic, the optimized configuration is usually obtained in a fixed boundary[28], and the limitations on vacuum fields are not always clear. Moreover, fixed-boundary steady states could be also used to examine the robustness of the designed stellarator configuration[29].

In this paper, we present a newly developed solver, called the Non-axisymmetric Toroidal Equilibrium Code (NTEC), that has incorporated all the above desired features. In particularly, NTEC finds a resistive MHD equilibrium that satisfies the force balance equation $\boldsymbol{J} \times \boldsymbol{B} = \nabla p$, by introducing an artificial flow to dynamically relax the magnetic field, and using a viscous damping[30] to maintain the $\boldsymbol{B} \cdot \nabla p = 0$ condition, which avoids additional assumptions on the magnetic field itself. Furthermore, NTEC employs the spectral/finite-element method to solve the set of nonlinear equations utilized in the iteration process, where the poloidal plane is discretized using the full Bell triangle[31] with first-order derivative continuity (i.e. C1-continuity) and quintic basis functions, and the grids in the toroidal direction are represented using finite Fourier series. Therefore, the discontinuous jump or spikes in current density and residual force, carried from the discontinuity of the first or second derivatives of magnetic fields, can be significantly mitigated due to the C1-continuity. In the meantime, the C1-continuity makes it possible to directly invert a second-order linear operator, which can be greatly beneficial for implementation if a physical quantity is cast in its potential representation. In addition, high-order numerical methods provide a favorable extensibility. Although a second-order accurate method is usually sufficient for static MHD equilibrium solutions, high-order methods are always preferred or even required in state-of-art MHD simulations of plasma evolution, for example the JOREK/JOREK3D[32], M3D-C1[33], and NIMROD/NIMSTELL[34] codes, among others. Compared to low-order methods, we need not introduce auxiliary variables when dealing with high-order physical derivative involved in the dissipation terms, such as viscosity, resistivity, and even hyper-diffusivity. The high-order method implemented in NTEC allows us to include more similar effects in future extensions.

The remainder of this paper is organized as follows. Section 2 elaborates on the numerical schemes including the discretization method, coordinate mapping, and equations for the iterative solving procedure. Section 3 describes the equilibria in stellarator geometry. Section 4 demonstrates the cases for helical equilibria in tokamak and QSH state in RFP. Finally, Section 5 presents the summary and further work.

## 2. Numerical schemes

### 2.1 C1-continuous quintic triangular element and Fourier decomposition

For toroidal systems, NTEC naturally adopts a finite Fourier series to represent quantities in the toroidal direction, and utilizes the Bell triangular element[31] in the poloidal plane. In order to illustrate the Bell element, consider a quantity expressed as a general polynomial:

$$F(x, y) = \sum_{j=1}^{N} b_j x^{m_j} y^{n_j},$$

where $N$ is the number of terms, $b_j$ is the coefficient, $m_j$ and $n_j$ are the exponents.

The $P_K$ Bell triangular element ($P_K$ denotes the space of piecewise polynomials and $K$ is the order), generally speaking, refers to the one that has no edge-degree of freedom and contains $P_{K-1}$ space locally[35]. For example, $\partial_n F$ will be a cubic polynomial function along each edge if $F$ is a quintic function. The most used type is quintic, as higher-order Bell elements would introduce extra degrees of freedom that may have to be determined using $F$ values at interior points in triangles[36]. As for the quintic Bell triangle, where $N = 21$, three constraints with one on each edge are imposed to drop the quartic terms in the normal derivative so that $\partial_n F$ along the edge can be completely determined by $\partial_n F$ and $\partial_{nn} F$ of the two adjacent vertices, therefore, the C1-continuity is achieved. The remaining 18 constraints are determined by $F$, $\partial_x F$, $\partial_y F$, $\partial_{xx} F$, $\partial_{xy} F$, $\partial_{yy} F$ at three vertices. In the community of fusion plasma modeling, the M3D-C1 code first employs the Bell element[37]. It rotates triangles to a zero-azimuthal position to simplify 2D integrals of polynomials over a triangle, and thus, the number of terms $N$ reduces to 20 from 21. We present here the full Bell element without this manipulation.

To derive the transformation between the quantities $F$, $\partial_x F$, $\partial_y F$, $\partial_{xx} F$, $\partial_{xy} F$, $\partial_{yy} F$ at three points and 21 coefficients $b_j$, we define auxiliary coordinate variables $(u, v)$ for the affine transformation

$$\begin{cases} u = x - x_0 \\ v = y - y_0 \end{cases},$$

where $(x_0, y_0)$ is the barycenter of each triangle. Hence, the transformation matrix $T_{(21 \times 21)}$ that satisfies the relation $f_{(21 \times 1)} = T_{(21 \times 21)} \cdot b_{(21 \times 1)}$ can be compactly expressed as:

$$T = \begin{bmatrix} T_{1(6\times21)} \\ T_{2(6\times21)} \\ T_{3(6\times21)} \\ & R_{1(1\times6)} \\ 0_{(3\times15)} & R_{2(1\times6)} \\ & R_{3(1\times6)} \end{bmatrix},$$

where

$$f = \begin{bmatrix} F_1 & F_{1x} & F_{1y} & F_{1xx} & F_{1xy} & F_{1yy} & F_2 & F_{2x} & F_{2y} & F_{2xx} & F_{2xy} & F_{2yy} & F_3 & F_{3x} & F_{3y} & F_{3xx} & F_{3xy} & F_{3yy} & 0 & 0 & 0 \end{bmatrix}^T$$

$$T_i = \begin{bmatrix} 1 & u_i & v_i & u_i^2 & u_i v_i & v_i^2 & u_i^3 & u_i^2 v_i & u_i v_i^2 & v_i^3 & u_i^4 & u_i^3 v_i & u_i^2 v_i^2 & u_i v_i^3 & v_i^4 & u_i^5 & u_i^4 v_i & u_i 3 v_i^2 & u_i^2 v_i^3 & u_i v_i^4 & v_i^5 \\ 0 & 1 & 0 & 2u_i & v_i & 0 & 3u_i^2 & 2u_i v_i & v_i^2 & 0 & 4u_i^3 & 3u_i^2 v_i & 2u_i v_i^2 & v_i^3 & 0 & 5u_i^4 & 4u_i^3 v_i & 3u_i^2 v_i^2 & 2u_i v_i^3 & v_i^4 & 0 \\ 0 & 0 & 1 & 0 & u_i & 2v_i & 0 & u_i^2 & 2u_i v_i & 3v_i^2 & 0 & u_i^3 & 2u_i^2 v_i & 3u_i v_i^2 & 4v_i^3 & 0 & u_i^4 & 2u_i^3 v_i & 3u_i^2 v_i^2 & 4u_i v_i^3 & 5v_i^4 \\ 0 & 0 & 0 & 2 & 0 & 0 & 6u_i & 2v_i & 0 & 0 & 12u_i^2 & 6u_i v_i & 2v_i^2 & 0 & 0 & 20u_i^3 & 12u_i^2 v_i & 6u_i v_i^2 & 2v_i^3 & 0 & 0 \\ 0 & 0 & 0 & 0 & 1 & 0 & 0 & 2u_i & 2v_i & 0 & 0 & 3u_i^2 & 4u_i v_i & 3v_i^2 & 0 & 0 & 4u_i^3 & 6u_i^2 v_i & 6u_i v_i^2 & 4v_i^3 & 0 \\ 0 & 0 & 0 & 0 & 0 & 2 & 0 & 0 & 2u_i & 6v_i & 0 & 0 & 2u_i^2 & 6u_i v_i & 12v_i^2 & 0 & 0 & 2u_i^3 & 6u_i^2 v_i & 12u_i v_i^2 & 20v_i^3 \end{bmatrix},$$

$$R_i = \begin{bmatrix} +5 \sin \theta_i \, (u_i - u_{i+1})^4 \\ -(u_i - u_{i+1})^3 (u_i \cos \theta_i - u_{i+1} \cos \theta_i - 4v_i \sin \theta_i + 4v_{i+1} \sin \theta_i) \\ -(u_i - u_{i+1})^2 (v_i - v_{i+1})(2u_i \cos \theta_i - 2u_{i+1} \cos \theta_i - 3v_i \sin \theta_i + 3v_{i+1} \sin \theta_i) \\ -(u_i - u_{i+1})(v_i - v_{i+1})^2 (3u_i \cos \theta_i - 3u_{i+1} \cos \theta_i - 2v_i \sin \theta_i + 2v_{i+1} \sin \theta_i) \\ -(v_i - v_{i+1})^3 (4u_i \cos \theta_i - 4u_{i+1} \cos \theta_i - v_i \sin \theta_i + v_{i+1} \sin \theta_i) \\ -5 \cos \theta_i \, (v_i - v_{i+1})^4 \end{bmatrix}^T.$$

The loop index $i = 1, 2, 3$ denotes the three vertices and $\theta$ is the azimuthal angle of $(x, y)$. The inverse transformation matrix $G_{(21 \times 18)}$ is the first 18 columns of the inverse matrix of $T_{(21 \times 21)}$. Consequently, any scalar function in $x - y$ plane can be written as

$$F(x, y) = \sum_{j=1}^{21} b_j (x - x_0)^{m_j} (y - y_0)^{n_j} = \sum_{j=1}^{21} \left( \sum_{k=1}^{18} G_{j,k} F_k \right) (x - x_0)^{m_j} (y - y_0)^{n_j} = \sum_{k=1}^{18} v_k F_k,$$

where the 18 2D basis functions are

$$v_k = \sum_{j=1}^{21} G_{j,k} (x - x_0)^{m_j} (y - y_0)^{n_j}.$$

Fig. 1 shows the contour plots of six basis functions at one vertex.

The next step is to establish the formula for the 2D integral of a monomial over an arbitrary triangle. We directly give the result here since the detailed derivation is sort of cumbersome. By introducing the barycentric coordinates to simplify the computation, one has

$$\iint_{tri} x^m y^n dS = \frac{2Am!n!}{(2+m+n)!} \sum_{\substack{m_1+m_2+m_3=m \\ m_1,m_2,m_3\geq 0}} \sum_{\substack{n_1+n_2+n_3=n \\ n_1,n_2,n_3\geq 0}} \prod_{i=1}^{3} \left[ \frac{(m_i+n_i)!}{m_i!n_i!} x_i^{m_i} y_i^{n_i} \right],$$

where $(x_i, y_i)$ are the vertex coordinates and $A$ is the area of the relevant triangle. This formula can be checked using Gaussian quadratures.

Finally, a toroidal angle $\zeta$ is used to discretize the third dimension with a certain period. Now, a scalar quantity can be expressed as

$$F(x,y,\zeta) = \sum_{k=1}^{18} v_k(x,y) F_k(\zeta) = \sum_{k=1}^{18} \left\{ v_k(x,y) \sum_{n=1}^{N} \left[ \begin{matrix} F_{k,n}^c \cos\left(N_{fp} \frac{(n_{odd}-1)\zeta}{2}\right) + \\ F_{k,n}^s \sin\left(N_{fp} \frac{n_{even}\zeta}{2}\right) \end{matrix} \right] \right\} = \sum_{j=1}^{18N} \lambda_j \mathcal{F}_j,$$

where $N$ and $N_{fp}$ are the numbers of Fourier components and toroidal period, respectively. The 3D basis function is

$$\lambda_j = \sum_{i=1}^{21} \mathcal{G}_{i,j} (x-x_0)^{m_i} (y-y_0)^{n_i},$$

and the kernel $\mathcal{G}$ is defined as

$$\mathcal{G}_{i,j} = \begin{cases} G_{i,(j/N)}, & n=1 \\ \sin\left(N_{fp}\frac{n\zeta}{2}\right) G_{i,(j/N)}, & n=\text{even} \\ \cos\left(N_{fp}\frac{(n-1)\zeta}{2}\right) G_{i,(j/N)}, & n=\text{odd} \end{cases}.$$

2.2 Coordinates mapping

The spectral/finite-element approach described above is constructed within the axisymmetric grids and will be sufficiently valid for cases with axisymmetric boundary if we directly triangulate the physical poloidal plane. However, the capability of properly handling a non-axisymmetric mesh is crucial for stellarators. For this reason, it may be wiser to triangulate on a computational axisymmetric mesh, and set up a coordinates mapping from the computational domain $(x,y,\zeta)$ to the physical domain $(R,Z,\varphi)$. Nevertheless, it is quite complicated to construct a general mapping due to non-convex boundary shape in modern stellarators. The Schwarz–Christoffel conformal mapping [38] or bijective composite mapping [39] can tackle the non-convex difficulty, but may also undermine the toroidal period of grids. An effective method is to exploit the geometry data of nested flux surfaces and build a flux-surface-aligned mesh, which is called pseudo flux mapping here. The "pseudo flux" means we need not require the nested surfaces to exactly match the actual flux surfaces, but only the flux values of which to be in a monotonic sequence. It should be mentioned that pseudo flux mapping itself does not ensure the C1-continuity from a computational domain to the physical domain. As a result, we must use the C1-continuous basis functions in NTEC for its representation. For simplicity, we choose $\zeta = \varphi$, thus

$$R = R(x,y,\zeta), \quad Z = Z(x,y,\zeta), \quad \varphi = \zeta.$$

The representations of nested flux surfaces in ideal equilibrium codes are different. As in the VMEC code, the geometry here is described as $R(\psi_N, \theta, \varphi)$ and $Z(\psi_N, \theta, \varphi)$ with $\psi_N$ being the normalized toroidal or poloidal flux and $\theta$ being the poloidal angle. A convenient and useful choice is

$$\psi_N = x^2 + y^2, \quad \theta = \begin{cases} \arctan(y/x), & x > 0 \\ \arctan(y/x) + \pi, & y \geq 0, x < 0 \\ \arctan(y/x) - \pi, & y < 0, x < 0 \\ +\pi/2, & y > 0, x = 0 \\ -\pi/2, & y < 0, x = 0 \end{cases}.$$

Fig. 2 depicts the mesh transformation via the pseudo flux mapping, which connects the computational grids to the physical grids for the standard magnetic configuration of the quasi-axisymmetric stellarator CFQS[40].

After constructing such a coordinate mapping for the stellarator geometry, we can eventually treat properly the derivatives up to the second order in the physical domain. The transformation between the derivatives of the first two orders in the computational and the physical domains can be obtained as follows

$$\begin{bmatrix} F_R \\ F_Z \\ F_{RR} \\ F_{RZ} \\ F_{ZZ} \\ F_{\varphi R} \\ F_{\varphi Z} \\ F_{\varphi\varphi} \\ F_\varphi \end{bmatrix} = \begin{bmatrix} \frac{Z_y}{D} & -\frac{Z_x}{D} & 0 & 0 & 0 & 0 & 0 & 0 & 0 \\ -\frac{R_y}{D} & \frac{R_x}{D} & 0 & 0 & 0 & 0 & 0 & 0 & 0 \\ \frac{A_{31}}{D^3} & \frac{A_{32}}{D^3} & \frac{Z_y^2}{D^2} & -\frac{2Z_xZ_y}{D^2} & \frac{Z_x^2}{D^2} & 0 & 0 & 0 & 0 \\ \frac{A_{41}}{D^3} & \frac{A_{42}}{D^3} & -\frac{R_yZ_y}{D^2} & \frac{R_xZ_y + R_yZ_x}{D^2} & -\frac{R_xZ_x}{D^2} & 0 & 0 & 0 & 0 \\ \frac{A_{51}}{D^3} & \frac{A_{52}}{D^3} & \frac{R_y^2}{D^2} & -\frac{2R_xR_y}{D^2} & \frac{R_x^2}{D^2} & 0 & 0 & 0 & 0 \\ \Sigma\alpha_iP_{i1} & \Sigma\alpha_iP_{i2} & \Sigma\alpha_iP_{i3} & \Sigma\alpha_iP_{i4} & \Sigma\alpha_iP_{i5} & \frac{Z_y}{D} & -\frac{Z_x}{D} & 0 & 0 \\ \Sigma\beta_iP_{i1} & \Sigma\beta_iP_{i2} & \Sigma\beta_iP_{i3} & \Sigma\beta_iP_{i4} & \Sigma\beta_iP_{i5} & -\frac{R_y}{D} & \frac{R_x}{D} & 0 & 0 \\ \Sigma\gamma_iP_{i1} & \Sigma\gamma_iP_{i2} & \Sigma\gamma_iP_{i3} & \Sigma\gamma_iP_{i4} & \Sigma\gamma_iP_{i5} & \frac{2E}{D} & \frac{2H}{D} & 1 & 0 \\ \frac{E}{D} & \frac{H}{D} & 0 & 0 & 0 & 0 & 0 & 0 & 1 \end{bmatrix} \begin{bmatrix} F_x \\ F_y \\ F_{xx} \\ F_{xy} \\ F_{yy} \\ F_{\zeta x} \\ F_{\zeta y} \\ F_{\zeta\zeta} \\ F_\zeta \end{bmatrix},$$

where

$$\begin{cases} D = R_xZ_y - R_yZ_x \\ E = R_yZ_\zeta - R_\zeta Z_y, \\ H = R_\zeta Z_x - R_xZ_\zeta \end{cases} \quad \begin{cases} \boldsymbol{\alpha}_{(1\times5)} = [R_{\zeta y}Z_x - R_{\zeta x}Z_y \quad Z_{\zeta y}Z_x - Z_{\zeta x}Z_y \quad -DR_\zeta \quad -DZ_\zeta \quad 0]/D \\ \boldsymbol{\beta}_{(1\times5)} = [R_{\zeta x}R_y - R_{\zeta y}R_x \quad Z_{\zeta x}R_y - Z_{\zeta y}R_x \quad 0 \quad -DR_\zeta \quad -DZ_\zeta]/D \\ \boldsymbol{\gamma}_{(1\times5)} = [-2ER_{\zeta x} - 2HR_{\zeta y} - DR_{\zeta\zeta} \quad -2EZ_{\zeta x} - 2HZ_{\zeta y} - DZ_{\zeta\zeta} \quad DR_\zeta^2 \quad 2DR_\zeta Z_\zeta \quad DZ_\zeta^2]/D \end{cases},$$

$$\begin{cases} A_{31} = -R_{yy}Z_x^2Z_y + R_yZ_{yy}Z_x^2 + 2R_{xy}Z_xZ_y^2 - 2R_yZ_{xy}Z_xZ_y - R_{xx}Z_y^3 + R_yZ_{xx}Z_y^2 \\ A_{32} = R_{yy}Z_x^3 - 2R_{xy}Z_x^2Z_y - R_xZ_{yy}Z_x^2 + R_{xx}Z_xZ_y^2 + 2R_xZ_{xy}Z_xZ_y - R_xZ_{xx}Z_y^2 \\ A_{41} = R_{xx}R_yZ_y^2 - R_xR_{xy}Z_y^2 + R_y^2Z_xZ_{xy} - R_y^2Z_{xx}Z_y - R_xR_yZ_xZ_{yy} + R_xR_yZ_{xy}Z_y + R_xR_{yy}Z_xZ_y - R_{xy}R_yZ_xZ_y \\ A_{42} = R_{xy}R_yZ_x^2 - R_xR_{yy}Z_x^2 + R_x^2Z_xZ_{yy} - R_x^2Z_{xy}Z_y + R_xR_{xy}Z_xZ_y - R_xR_yZ_xZ_{xy} + R_xR_yZ_{xx}Z_y - R_{xx}R_yZ_xZ_y \\ A_{51} = Z_{yy}R_x^2R_y - R_{yy}Z_xR_x^2 - 2Z_{xy}R_xR_y^2 + 2R_{xy}Z_yR_xR_y + Z_{xx}R_y^3 - R_{xx}Z_yR_y^2 \\ A_{52} = -Z_{yy}R_x^3 + 2Z_{xy}R_x^2R_y + R_{yy}Z_xR_x^2 - Z_{xx}R_xR_y^2 - 2R_{xy}Z_xR_xR_y + R_{xx}Z_xR_y^2 \end{cases}.$$

### 2.3 Iteration process

Without the NFS assumption, the inverse representation of quantities via global flux coordinates is no longer always guaranteed. The solutions of the force balance equation $\nabla p = \boldsymbol{J} \times \boldsymbol{B}$, in the stochastic regions and island chains, could be pathological for numerical calculations[41]. The chaotic field lines within their vicinity could come arbitrarily close to each other, a local trivial solution thereby often seems attractive for the pressure field. However, there is no clear criterion distinguishing among the large islands, small island chains, and chaotic fields. In another word, the pressure should be "self-organized" in the iteration process of finding its entire numerical solution.

The key issue is how to take the redistribution of pressure into account reasonably. The adiabatic energy equation $\partial_t p = (\gamma - 1)\boldsymbol{u} \cdot \nabla p - \gamma \nabla \cdot (p\boldsymbol{u})$ does not directly seek solutions that satisfy $\boldsymbol{B} \cdot \nabla p = 0$. Thus, it is necessary to introduce a new formula out of the physical MHD equations for the pressure updating. Note that a parallel diffusion equation $\partial_t p = \nabla \cdot (D_\parallel \nabla_\parallel p)$ exactly flattens the pressure along magnetic field line. But this equation is still unfavorable for numerical implementation because the CFL condition requires the time step $\delta t$ to be $O(h^2)$ ($h$ is the grid size) in an explicit method while a fully implicit method may be too expensive for an equilibrium solver. A modification can be made through the viscous damping formulation as follows

$$\frac{\partial^2 p}{\partial t^2} + \frac{1}{\tau}\frac{\partial p}{\partial t} = \mathcal{L}(p) = \nabla \cdot (\nabla_\parallel p),$$

where the inverse damping factor $\tau$ controls the stability and convergence rate. For a nearly critical damping, we simply choose $\tau(R, Z, \varphi) \approx \sqrt{\mathcal{L}/p}$ and set $\tau_{max} \sim O(10)$ by default to avoid $\tau \to \infty$ when $p \to 0$. The hyperbolic equation allows us to take $\delta t \sim O(h)$, which significantly improves the convergence speed.

The approach to the magnetic field relaxation is straightforward. To consider the resistive effect in a non-variational method, the magnetic induction equation should remain as

$$\frac{\partial \boldsymbol{B}}{\partial t} = \nabla \times (\boldsymbol{u} \times \boldsymbol{B} - \eta_0 \boldsymbol{J}).$$

For the calculations where the potential representation is not used, an artificial diffusion term $\kappa_{\text{div}} \nabla \nabla \cdot \boldsymbol{B}$ is added to comply with the divergence-free condition, and $\kappa_{\text{div}}$ is set to be a large number (the value between $10^4$ and $10^6$ is generally sufficient). In this dynamic relaxation process of the magnetic field, the artificial plasma flow cannot be omitted. Otherwise, any initial equilibrium tends to approach a vacuum solution when $t \to \infty$. We take the perpendicular component $\boldsymbol{u}_\perp \approx (\boldsymbol{B} \times \boldsymbol{J})/B^2$ as an initial flow to obtain a moderately perturbed state after several iteration steps, which is beneficial for reducing the numerical oscillations in high $\langle \beta \rangle$ regimes ($\langle \beta \rangle$ is the volume-averaged ratio of the plasma pressure $p$ to the magnetic pressure $B^2/2\mu_0$).

The artificial flow is no doubt accumulated due to the residual force,

$$\rho \frac{\partial \boldsymbol{u}}{\partial t} = \boldsymbol{J} \times \boldsymbol{B} - \nabla p + \nu \nabla^2 \boldsymbol{u}.$$

The plasma density $\rho$ has little influence on the force balance condition in the final steady state, and thus is set to be a constant. Additionally, an optimal choice[42] for $\rho = (B^2/2\mu_0 + \gamma p)/v_{fast}^2$ with an arbitrary uniform magneto-sonic wave speed $v_{fast}$ has been tested, but we find it contributes little to the performance. A small viscosity term is used to keep the numerical stability, acquire a smooth flow, and also somehow serve as compensation for the stochastic regions that may inherently violate the static force balance owing to the opening magnetic field line[43]. The advection term is dropped, thereby the artificial flow is almost localized. It is expected that, from an ideal equilibrium or a vacuum solution, the iteration stops when $\boldsymbol{J} \times \boldsymbol{B} = \nabla p$ and $\partial_t \boldsymbol{B} = 0$.

In this work, the time advancing of the pressure and magnetic field are not synchronized. Their step intervals are entirely different, and at every step for the magnetic field, there is a sub-loop for the pressure update. All linear terms are calculated by using implicit methods while the nonlinear terms are explicitly handled, which allows larger time steps. The boundary conditions are no-slip for the velocity and fixed for the magnetic field. The resistivity, viscosity and volume integral of pressure $\iiint p dV$ are set to be constants. A total of $1454$ triangular elements are applied to all the NTEC calculations, with $25$ toroidal modes utilized for stellarator, and $15$ for tokamak and RFP.

## 3. Intrinsically 3D equilibria with non-axisymmetric boundary

### 3.1 Equilibrium benchmarks

It is very challenging to carry out a benchmark of resistive equilibria obtained from different solvers since a converged solution is sensitive to initial parameters and numerical schemes. The previous study and comparison[44,45] of the HINT and PIES free-boundary calculations for the large-volume W7-X configuration with $\langle\beta\rangle \approx 4\%$ showed visible distinctions in the view of Poincare plots. Even under the framework of ideal MHD incorporated with NFS assumption, a bifurcated 3D equilibrium that has almost the same potential energy against the 2D one could appear by merely disturbing the 2D magnetic axis[46]. Nevertheless, it is still essential to guarantee the reliability of a newly developed code before exhibiting any further calculation results. For this purpose, our primary objective in this part is, to some extent, reproducing the equilibria computed by other codes. We show the benchmarks of both fixed-boundary and free-boundary equilibria.

First, we display the results of a fixed-boundary low $\langle\beta\rangle$ case[20] on a classic stellarator where the SIESTA calculations are also performed. The fixed boundary for this stellarator with $N_{fp} = 3$ is given by

$$\begin{cases} R_b = 2.90 + \cos\theta - 0.51\cos(\theta - N_{fp}\varphi) - 0.01[\cos(4\theta + N_{fp}\varphi) + \cos(6\theta + N_{fp}\varphi)] \\ Z_b = \sin\theta + 0.51\sin(\theta - N_{fp}\varphi) + 0.01[\sin(4\theta + N_{fp}\varphi) + \sin(6\theta + N_{fp}\varphi)] \end{cases}.$$

Our calculations start with well-converged VMEC equilibria, which have used 500 flux surfaces in this work. In Ref. 20, the pressure profile of the initial ideal equilibrium calculated using VMEC is not explicitly described, and we use the profile $p = p_0(1 - \psi_N)^2$ to achieve a similar equilibrium with $\langle\beta\rangle \approx 10^{-4}$ (Here $\psi_N$ is the normalized toroidal flux in stellarator/tokamak, but poloidal flux in RFP). There is an expectation that, as the rotational transform $\iota$ goes from around $0.45$ to $0.74$, the islands will appear at three low-order rational values where $\iota = 3/6,\ 6/11,\ 3/5$. The artificial resistivity chosen in SIESTA is a fraction of the CFL value[20], and here we benchmark it against two values $1 \times 10^{-5}$ and $3 \times 10^{-5}$ in the unit of $\Omega \cdot m$ in the NTEC calculations.

We compare the Poincare plots for the equilibria calculated using the two codes as shown in Fig. 3. In the NTEC result with the relatively low resistivity $\eta = 1 \times 10^{-5}$, the $m = 5$ islands in the periphery emerge first while the internal islands are still less obvious. As the resistivity rises to be $3 \times 10^{-5}$, the NTEC result essentially coincides with the SIESTA one, where the inner $m = 6$ islands become obvious and $m = 5$ islands saturate. To demonstrate the convergence in detail, we show in Fig. 4 the variations of four metrics during the iteration in the case with $\eta = 3 \times 10^{-5}$. As to the first normalized quantity $\langle|\boldsymbol{B}\cdot\nabla p|/|\boldsymbol{B}|\rangle/\langle|\boldsymbol{B}_0\cdot\nabla p_0|/|\boldsymbol{B}_0|\rangle$ that measures the proximity to the $\boldsymbol{B}\cdot\nabla p = 0$ condition, it is no surprise it jumps up at the first few steps, because the pressure is a flux function only in VMEC where the ideal equilibrium strictly obeys the $\boldsymbol{B}\cdot\nabla p = 0$ condition, especially in an extremely low $\langle\beta\rangle$ case. It can be further reduced during the early iterations by increasing the number of sub-loops for pressure updating, but has little effect on equilibrium in such a low $\langle\beta\rangle$ scenario. The normalized residual force $\langle|\boldsymbol{F}_{MHD}|\rangle/\langle|\boldsymbol{F}_{MHD_0}|\rangle$ descends at first owing to the elimination of the singularity at the magnetic axis in the initial VMEC solution, and then increases due to the slow but cumulative build-up of chaotic fields causing $\boldsymbol{B}\cdot\nabla p \neq 0$. This process can be further revealed with the initial and final distributions of the normalized residual force as shown in Fig 5. The third row in Fig. 4 shows the change rate of the magnetic field strength, and the tolerance of $\langle\partial_t|\boldsymbol{B}|\rangle$ is set to be $1 \times 10^{-5}$ in both cases. The last quantity in Fig. 4 manifests the divergence-free property. Unless otherwise noted, all iteration ends at $\langle|\boldsymbol{F}_{MHD}|\rangle/\langle|\boldsymbol{F}_{MHD_0}|\rangle \approx 0.1\sim 1$ and $\partial_t\langle|\boldsymbol{B}|\rangle \approx 0.1\sim 5 \times 10^{-4}$.

Next, we compare the free-boundary NTEC calculation results for high $\langle\beta\rangle$ CFQS equilibria with the HINT calculation results[47]. We first elucidate the method of constructing a reasonable guess for the magnetic field in the whole computational volume. A direct interpolation of the VMEC field and the extended field, which here refers to the vacuum field produced by 16 modular coils[40] fails to provide a sufficiently smooth magnetic field. The second approach is to sum up the plasma-generating field using the so-called virtual casing principle[48] and the vacuum field via the Biot-Savart law. However, such a combined field still violates the divergence-free condition[49,50]. To resolve this problem, we use the magnetic vector potential to merge the two fields. The vector potential is obtained by inverting the double-curl operator under the Coulomb gauge and the essential boundary condition contributed from the plasma and coil current, which in turn gives the magnetic field.

Fig. 6 shows the Poincare plots of two free-boundary CFQS equilibria with $\langle \beta \rangle \approx 1.40\%$ and $1.58\%$ respectively, calculated using HINT code (also presented in Fig. 2 from Ref. 47). It can be seen that small islands with $n/m = 6/23$ appear in both equilibria and large islands with $n/m = 2/6$ occur only in the latter one. In the NTEC calculations, we start with a VMEC equilibrium with a proportionally enhanced pressure profile that $p \propto (1 - \psi_N)^2$, resulting in a higher $\langle \beta \rangle \approx 1.80\%$, as the plasma inevitably diffuses to the vacuum region and $\langle \beta \rangle$ thereby goes down in the iteration. The vacuum field is somewhat different, since we use a high-resolution filament model for each modular coil and the $n \neq 2$ symmetry-breaking components (which mainly reside near the coils) are ignored. Fig. 7 shows the Poincare plots of the equilibria calculated using NTEC with $\eta = 1.0 \times 10^{-5}$, $3.0 \times 10^{-5}$, and $1.0 \times 10^{-4}$ respectively. It is found that the resistivity has a significant influence on the magnetic configuration. As the resistivity grows, the radial location of the magnetic axis portrayed in Fig. 7 inwardly shifts from $0.927$ to $0.911$ to $0.889$, corresponding to a decreasing $\langle \beta \rangle$. For the highest resistivity equilibrium in which the $n/m = 6/23$ islands still persist, the $n/m = 2/6$ islands eventually disappear. In a degree, the variation trend of magnetic islands shows qualitative agreement with the HINT results. Additionally, the convergence metrics for the NTEC equilibrium with $\eta = 1.0 \times 10^{-4}$ is illustrated in Fig. 8.

3.2 Equilibria in a modern stellarator configuration

To produce comparable fixed-boundary CFQS equilibria with the free-boundary ones, we initialize the NTEC calculations with a $\langle \beta \rangle \approx 2.0\%$ VMEC equilibrium that keeps the pressure profile $p \propto (1 - \psi_N)^2$ whereby the rotational transform still passes through $1/3$. Three resistive equilibria, where one has a fixed pressure profile with $\eta = 1.0 \times 10^{-5}$, and two have self-organized pressure profiles with $\eta = 1.0 \times 10^{-5}$ and $1.0 \times 10^{-4}$ respectively, are considered.

The Poincare plots at two cross sections $\varphi = 0, \pi/2$ are shown in Fig. 9. For the fixed-pressure low-resistivity one, the rotational transform crosses over $1/3$ twice and therefore inner $n/m = 2/6$ islands and outer $n/m = 4/12$ islands emerge. The two resistive equilibria with adaptive pressure have distinct $n/m = 2/6$ islands. The low-resistivity case has similarly sized islands with the fixed-pressure one, which reveals that the $n/m = 4/12$ islands in the fixed-pressure equilibrium are indeed caused by the overlapping of inner and outer $n/m = 2/6$ saturated islands that have different poloidal phases. While in the high-resistivity case, the islands grow up evidently, and the magnetic field tends to become chaotic in the peripheral region. Fig. 10 shows the rotational transform profiles of the corresponding equilibria. The rotational transform $\iota$ near the plasma center has a drastic change compared with the ideal one, in both the low-resistivity fixed-pressure case and the high-resistivity adaptive-pressure case, likely due to the formation of the inner islands. As it can be seen that in the low-resistivity adaptive-pressure case the inner islands are nearly absent, and the change of rotational transform in the center remains relatively small.

## 4. Non-axisymmetric equilibria in nominally axisymmetric systems

### 4.1 Helical-core equilibrium states in tokamak plasmas

Bifurcated equilibria with a large helical core in tokamaks have been observed numerically by means of VMEC/ANIMEC[46]. Further VMEC calculations expose two types of helical states caused by current-driven internal-kink or pressure-driven quasi-interchange instability[51]. However, all the equilibria have an almost central shear-free safety factor profile with $q_0$ close to unity, and thus may evolve to new steady states that have a dominant inner island in the resistive model.

The resistive calculations are initialized with the VMEC equilibrium presented in Ref. 46 that has $p = p_0(1 - \psi_N)$, $\langle \beta \rangle \approx 0.5\%$, $\iota = 0.9 + 0.2\psi_N - 0.8\psi_N^6$ that corresponds to the safety factor profile with $q \approx 1$ from $\psi_N = 0$ to $0.64$ and $q \approx 3.3$ at the last closed flux surface, and a TCV-like plasma boundary as

$$\begin{cases} R_b = 0.8 + 0.2\cos\theta + 0.06\cos 2\theta \\ Z_b = 0.48\sin\theta \end{cases}.$$

We use two resistivities $\eta = 1.0 \times 10^{-5}$ and $1.0 \times 10^{-4}$ to obtain two bifurcated equilibria.

The comparisons among the initial ideal and two resistive equilibria using the Poincare plots and pressure contours in Fig. 11. show that, in the resistive solutions, a helical equilibrium with a large internal island is eventually reached as a result of the nonlinear saturation of the $n/m = 1/1$ resistive kink-tearing mode. The pressure profile is almost flattened within the main island and yet subtly differs in each case. In particular, the pressure magnitude within the $n/m = 1/1$ island is about half of $p_0$ in the low-resistivity case, while in the high-resistivity case the corresponding pressure has a comparable magnitude to $p_0$. The pressure distributions in the central region are distinct between the two resistive equilibria, and their influence on the magnetic configuration can be revealed by the safety factor profile as shown in Fig. 12. It is found that, in the low-resistivity case, the significant negative magnetic shear exactly coincides with the steep pressure gradient that extends from the vicinity of the magnetic axis to the separatrix of the internal saturated island. This local negative shear is beneficial for stabilizing the pressure-driven MHD modes[52], which leads to the well-converged solutions. On the other hand, the high-resistivity equilibrium keeps $q_0$ slightly above unity with low magnetic shear and a nearly flat pressure profile in the central area. The accessibility of this kind of equilibrium has been verified in the calculation results using QSOLVER and M3D-C1[53].

### 4.2 Quasi-single-helicity equilibria in a reversed-field pinch

Plasma current in RFPs is typically one order larger than that in tokamaks with a comparable size, leading to much more intense MHD activities. Especially in the configuration with reversed toroidal fields near the edge, the $m = 0$ resonant tearing modes often resolutely prevent plasmas from evolving into an equilibrium state. In order to find a converged solution, in the reversed-toroidal-field configuration, we subtract the ideal equilibrium current $\mathbf{J}_0$ in the dissipative term of the magnetic induction equation, which in a way serves as an effective source to sustain the final steady state[29].

In this section, we show a non-resonant QSH equilibrium with a slightly reversed toroidal field, and another resonant non-reversed QSH equilibrium of the KTX[54] plasmas that have major radius $R_0 = 1.4$ m and minor radius $r_0 = 0.4$ m. We take $N_{fp} = 6$ in these cases since previous QSH studies[55] in simulations and experiments confirm that the dominant helical mode occurs with $m = 1$ and $n \geq 3R_0/2r_0$. Regarding the initial safety factor profiles for VMEC calculations, we simply set $q_0 = Q_{6,7}$, $q_{0.5} = Q_{7,8}$, $q_1 = -0.02$, and $q_0 = 0.2$, $q_{0.5} = Q_{6,7}$, $q_1 = 0.07$ for the reversed and non-reversed equilibria, respectively. The value $Q_{a,b} =$

$(1 + \gamma)/(a + b\gamma)$ denotes the "most irrational" number between two low rational surfaces[56] in which $\gamma$ is the golden ratio. The initial pressure profile is $p = p_0(1 - \psi_N)^2$ with $\langle \beta \rangle \approx 1.0\%$ for the reversed one and $\langle \beta \rangle \approx 0.3\%$ for the other. The artificial resistivity is $1.0 \times 10^{-4}$.

The Poincare plots and pressure contours of both non-resonant and resonant KTX-QSH equilibria Fig. 13, along with the corresponding safety factor profiles in Fig. 14, show that the structure of the flux surfaces in the non-resonant case closely resembles the initial VMEC equilibrium, except for the radial locations associated with $q = 1/7$ and $1/8$, each of which has a high-order harmonic, i.e., $m/n = 6/42$ and $3/24$. Presumably, the nonlinear interactions among the three modes ($m/n = 1/6,7,8$) will lead to a significantly altered magnetic topology when dropping the constraint of toroidal periodicity and using a large number of Fourier modes. On the other side, the second ideal equilibrium converges to a resonant QSH equilibrium because of $q_0 > 1/6$ at the start. The flux surfaces resemble those encountered in TCV cases characterized by a strong shift of the magnetic axis, but in fact, they could be different. The magnetic axis may actually be embedded in the central stochastic region, whereas the primary magnetic island with hot plasmas winds around the center. Such a state bears a notable similarity to the double-axis-type QSH state already identified in experiments[57].

## 5. Summary and further work

This paper presents a spectral/finite-element code, NTEC, to numerically solve for the resistive finite-pressure MHD equilibria in non-axisymmetric toroidal systems. The quintic Bell triangles are used to discretize the poloidal plane, and the grids in periodic toroidal direction are represented using finite Fourier series. This combination guarantees the C1-continuity that reduces the error caused by the spikes or discontinuous jumping in quantities across element boundaries and allows for the handling of high-order operators. Furthermore, a pseudo flux mapping provides the capability for non-axisymmetric geometry as in stellarators. The pressure and magnetic field are iteratively updated using the hyperbolic parallel diffusion and dynamic relaxation equations, respectively. For the calculation results, we first benchmark the NTEC solutions with a fixed-boundary classic stellarator equilibrium from SIESTA and the free-boundary CFQS equilibria from HINT. And then we demonstrate the NTEC results in the fixed-boundary CFQS equilibria, the TCV-like helical-core equilibria, and the quasi-single-helicity equilibria in KTX.

As a newly developed code, several improvements could be implemented in future. The first effort is to test other types of Bell triangular elements. For example, one such type employs a stable B-spline representation that makes the Bell triangle non-negative for most points[58], which is beneficial for keeping positive values of pressure near the plasma-vacuum separatrix. Second, the boundary condition for the magnetic field needs a more rigorous treatment. It is convenient to use the ideal wall condition for now. However, if we use the induction equation to locally update the magnetic field on the boundary, the artificial flow gives rise to an anomalously accumulated magnetic field that may depart from a physical solution. It is suggested that we should consider the Nitsche's method[59] to weakly impose the boundary conditions, or alternatively update the magnetic field from the total plasma current using Biot-Savart law. The third direction is to enhance the numerical stability especially when using a large time step, and a generalized toroidal angle can be adopted to mitigate the rapid spatial variation of the Jacobian after the pseudo flux mapping. In brief, further work will include refined numerical techniques and focus on applications of NTEC to addressing 3D physical issues.

Acknowledgement

We sincerely thank Xianqu Wang (SWJTU) for sharing the data of HINT calculation and Haifeng Liu (SWJTU) for helpful discussions. This work was supported by the National Key Research and Development Program of China (Grant No. 2019YFE03050004), the National Natural Science Foundation of China (Grant No. 51821005), and the U.S. Department of Energy (Grant No. DE-FG02-86ER53218). The computing work in this paper was supported by the Public Service Platform of High Performance Computing by Network and Computing Center of HUST.

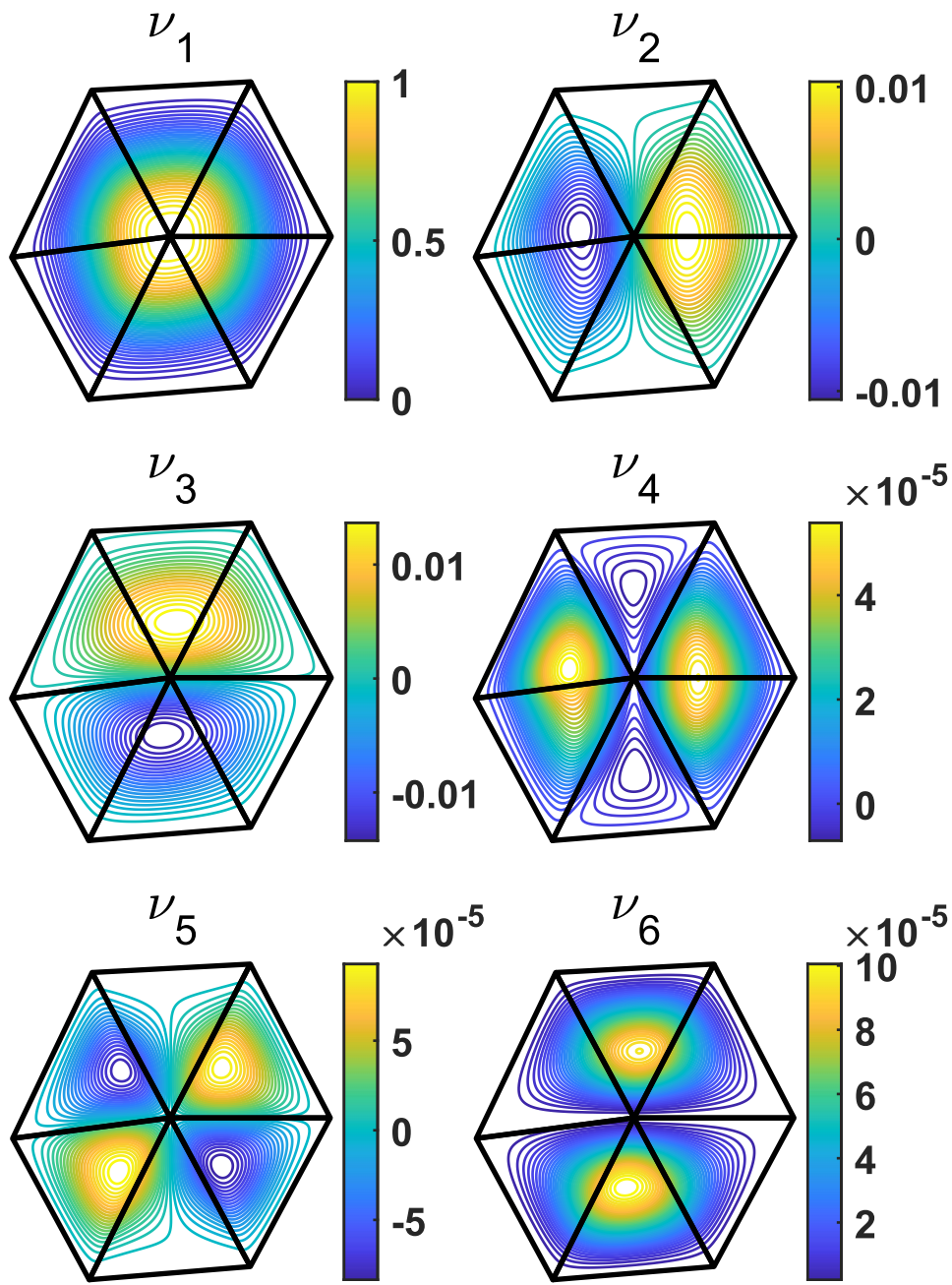

Fig. 1. Contour plots for 6 nodal basis functions of an interior vertex.

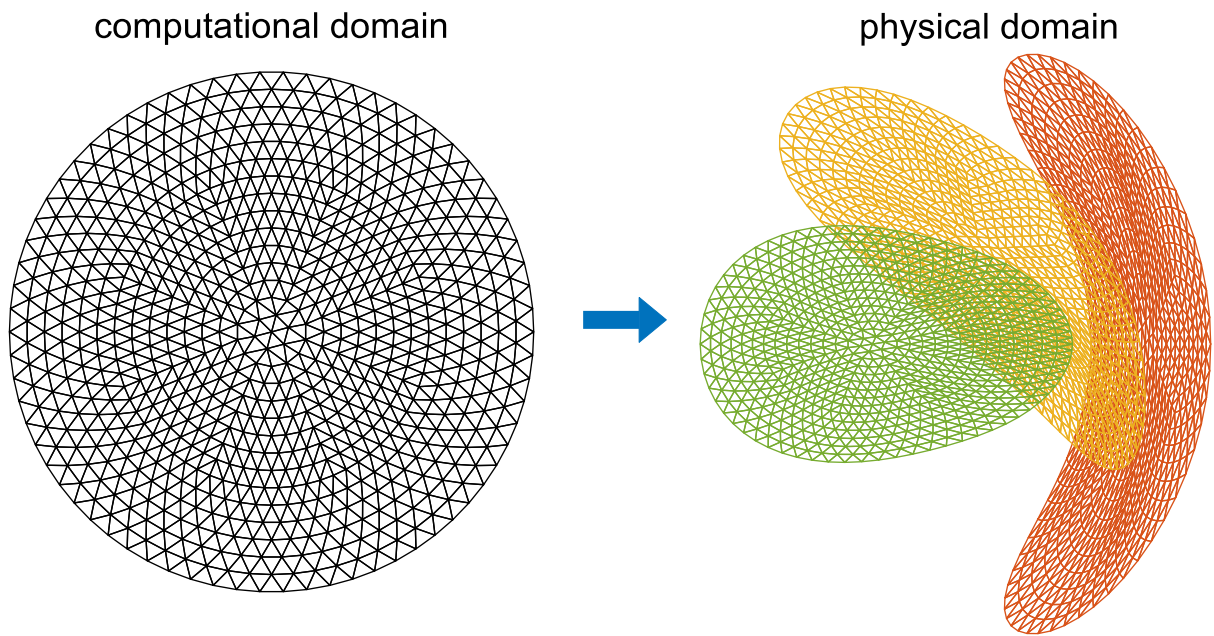

Fig. 2. Left: an axisymmetric mesh in the $x-y$ plane. Right: non-axisymmetric meshes in the $R-Z$ plane for the standard CFQS configuration, the red, yellow and green cross sections are at $\varphi = 0, \pi/4$, and $\pi/2$, respectively.

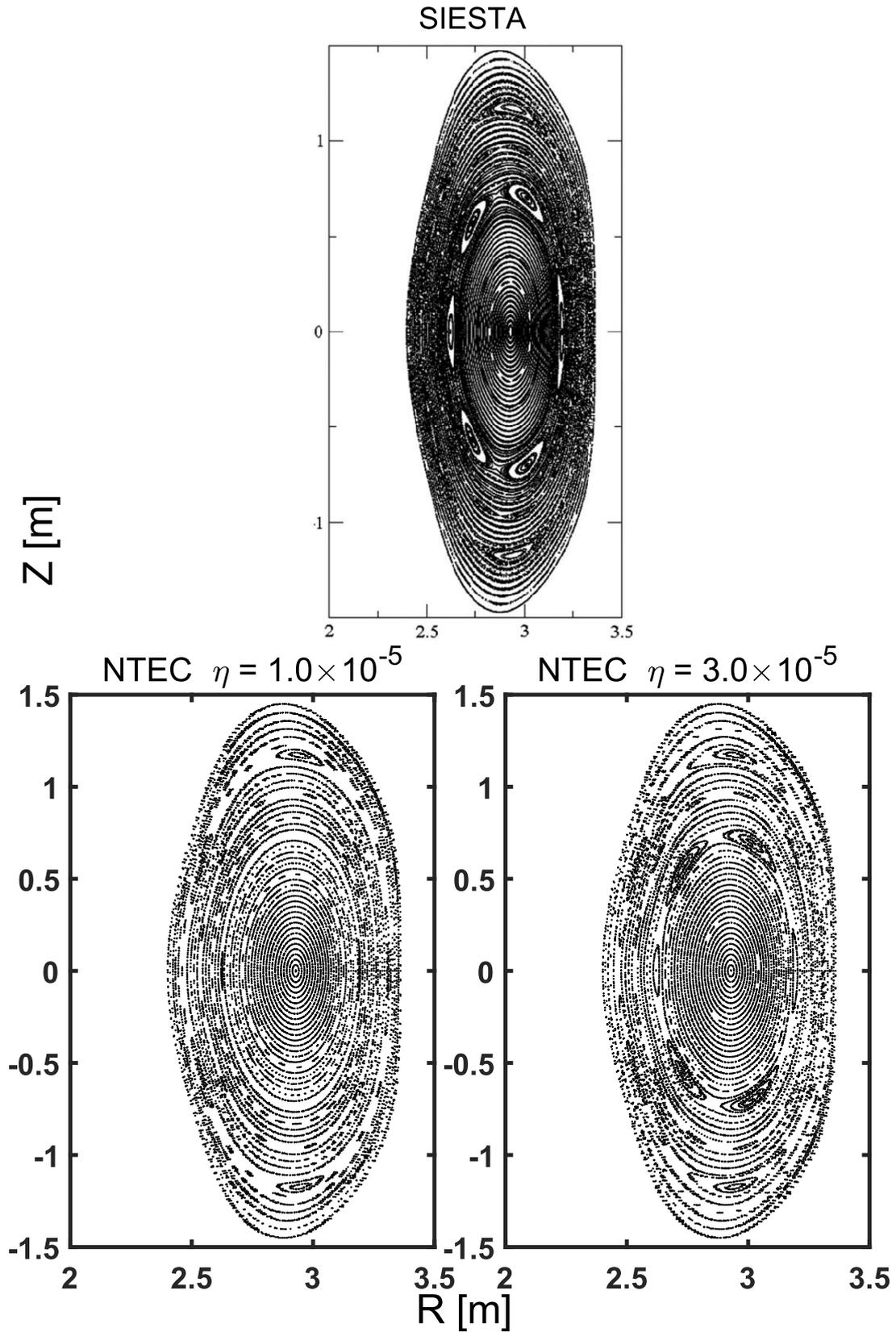

Fig. 3. Poincare plots in the $\varphi = 0$ cross section of the classic stellarator equilibria computed using (upper:) SIESTA (Reproduced with permission from Ref. 20, Copyright 2011, AIP Publishing); (lower left:) NTEC with $\eta = 1 \times 10^{-5}$; and (lower right:) NTEC with $\eta = 3 \times 10^{-5}$.

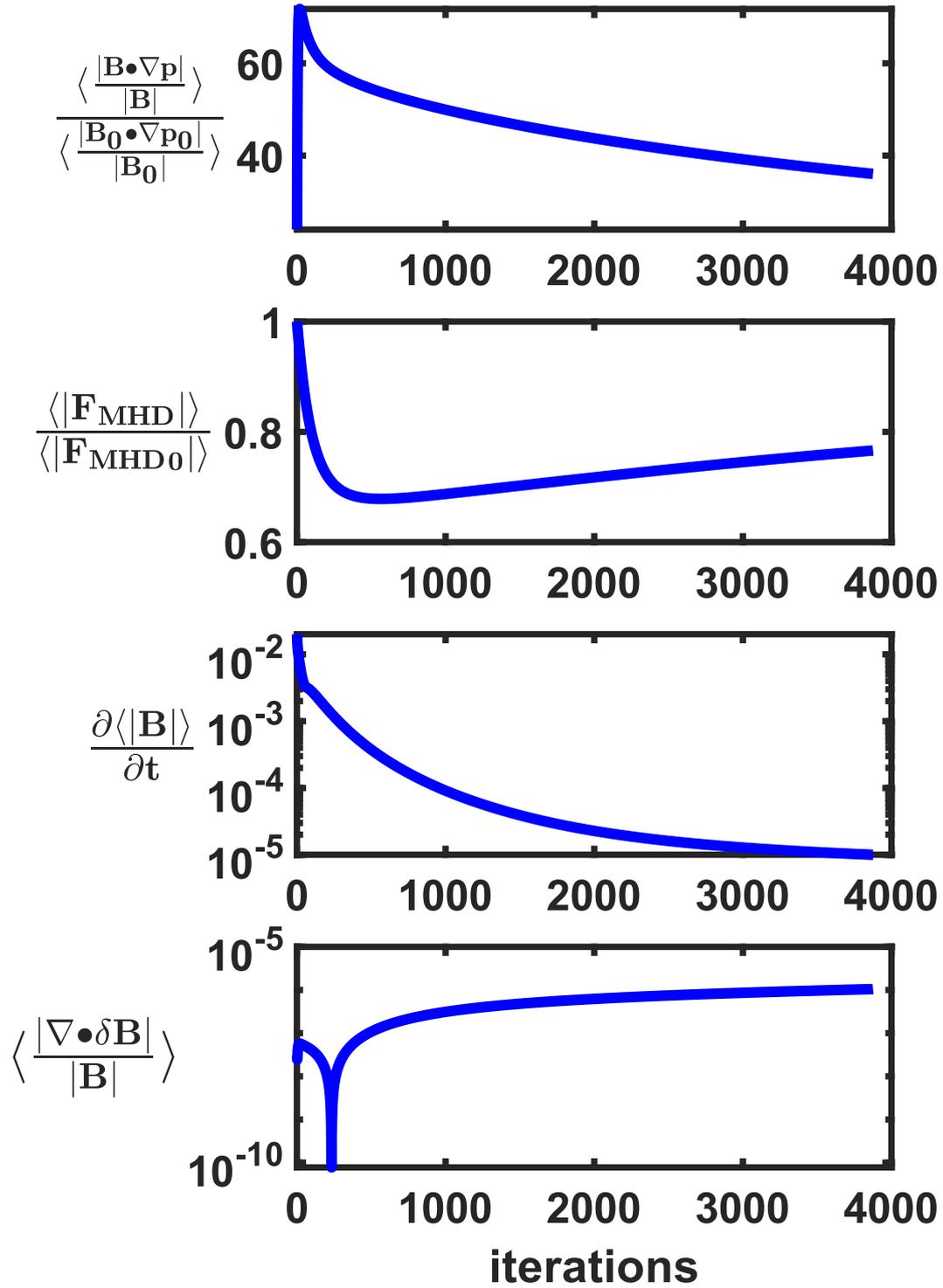

Fig. 4. Variations of the convergence metrics versus iteration steps for the classic stellarator case with $\eta = 3 \times 10^{-5}$. $\langle \cdots \rangle$ denotes the volume-averaging operator, the subscript "0" represents quantities from the initial equilibrium computed using VMEC.

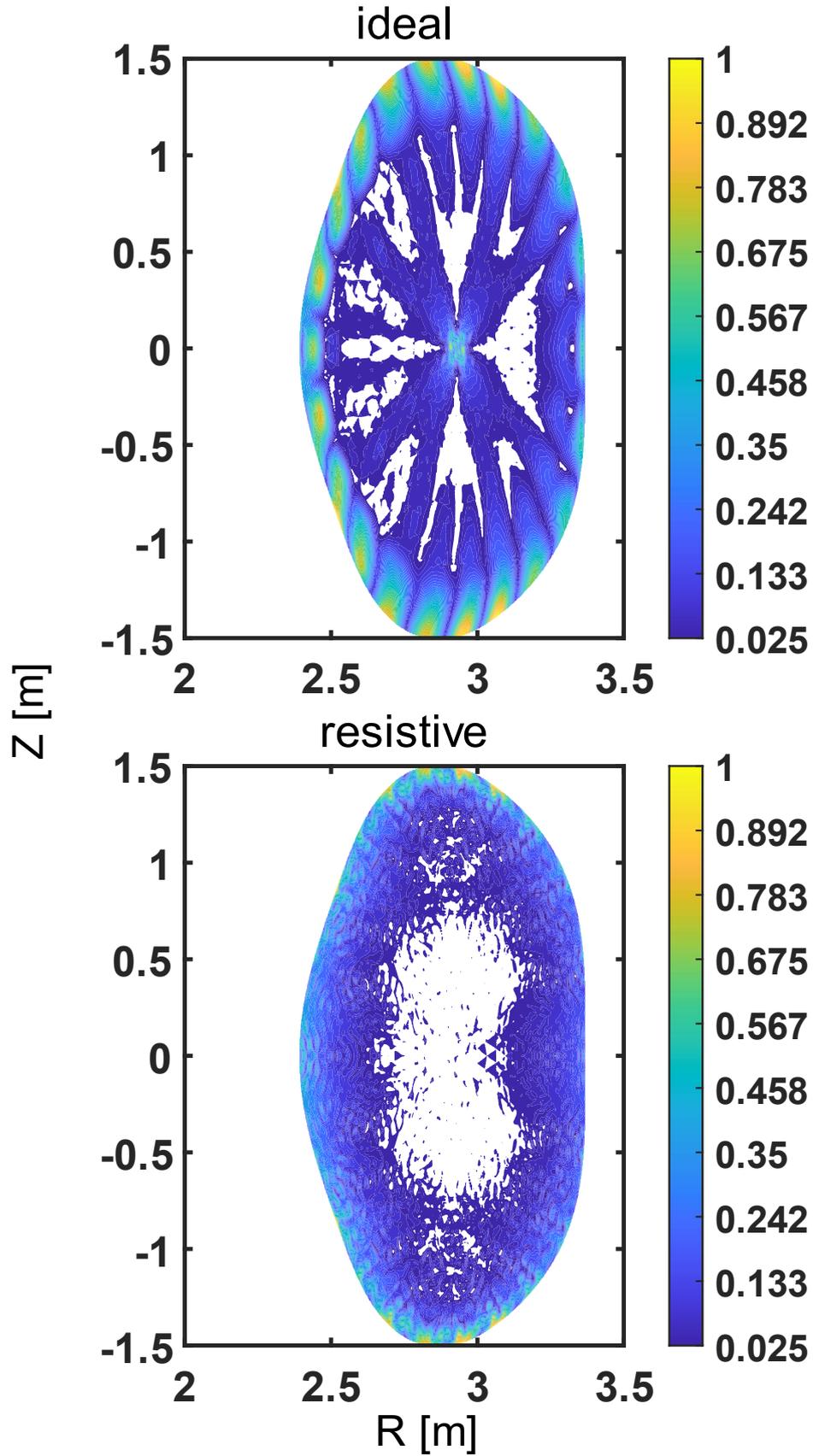

Fig. 5. Distributions of the residual forces normalized to the maximum value in the $\varphi = 0$ cross section for the solutions from (upper:) VMEC equilibrium; and (lower:) NTEC equilibrium with $\eta = 3 \times 10^{-5}$.

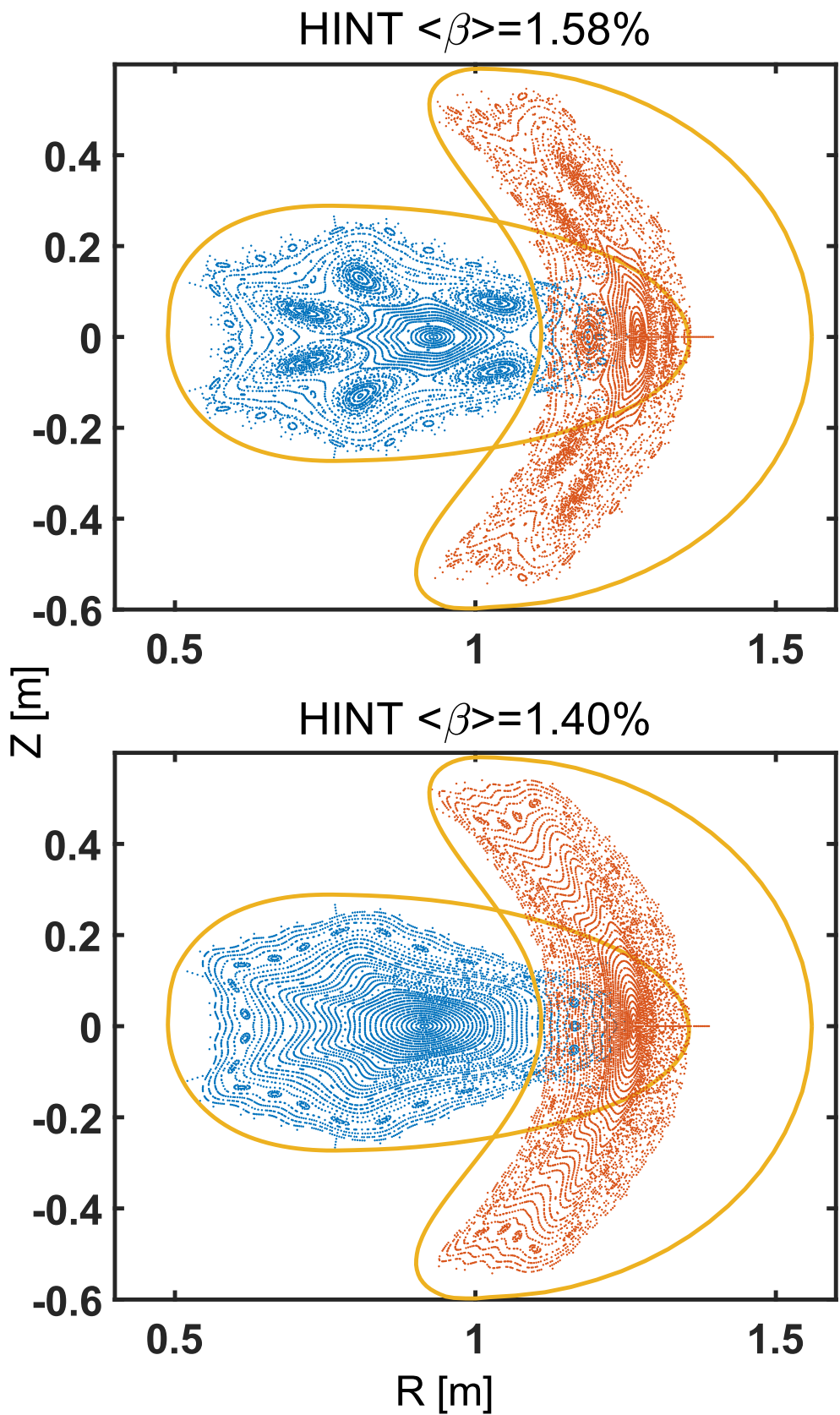

Fig. 6. Poincare plots of free-boundary CFQS equilibria from HINT for $\langle\beta\rangle = 1.58\%$ (upper) and $1.40\%$ (lower) at two cross sections $\varphi = 0$ (red), $\pi/2$ (blue). The yellow lines indicate the vacuum vessel boundaries.

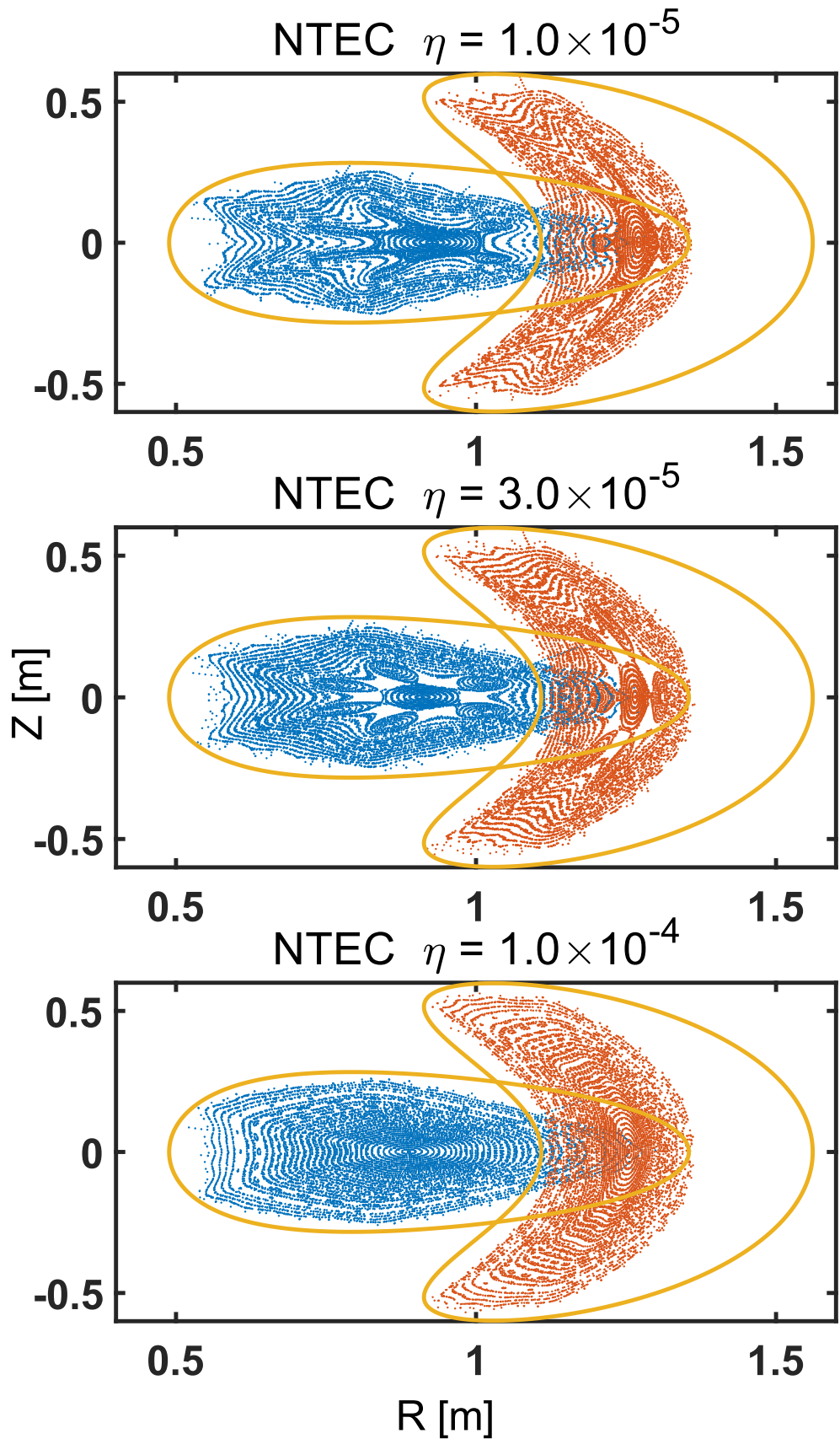

Fig. 7. Poincare plots of equilibria obtained from NTEC for $\eta = 1.0 \times 10^{-5}$ (top), $3.0 \times 10^{-5}$ (middle) and $1.0 \times 10^{-4}$ (bottom) at two cross sections $\varphi = 0$ (bottom), $\pi/2$ (blue).

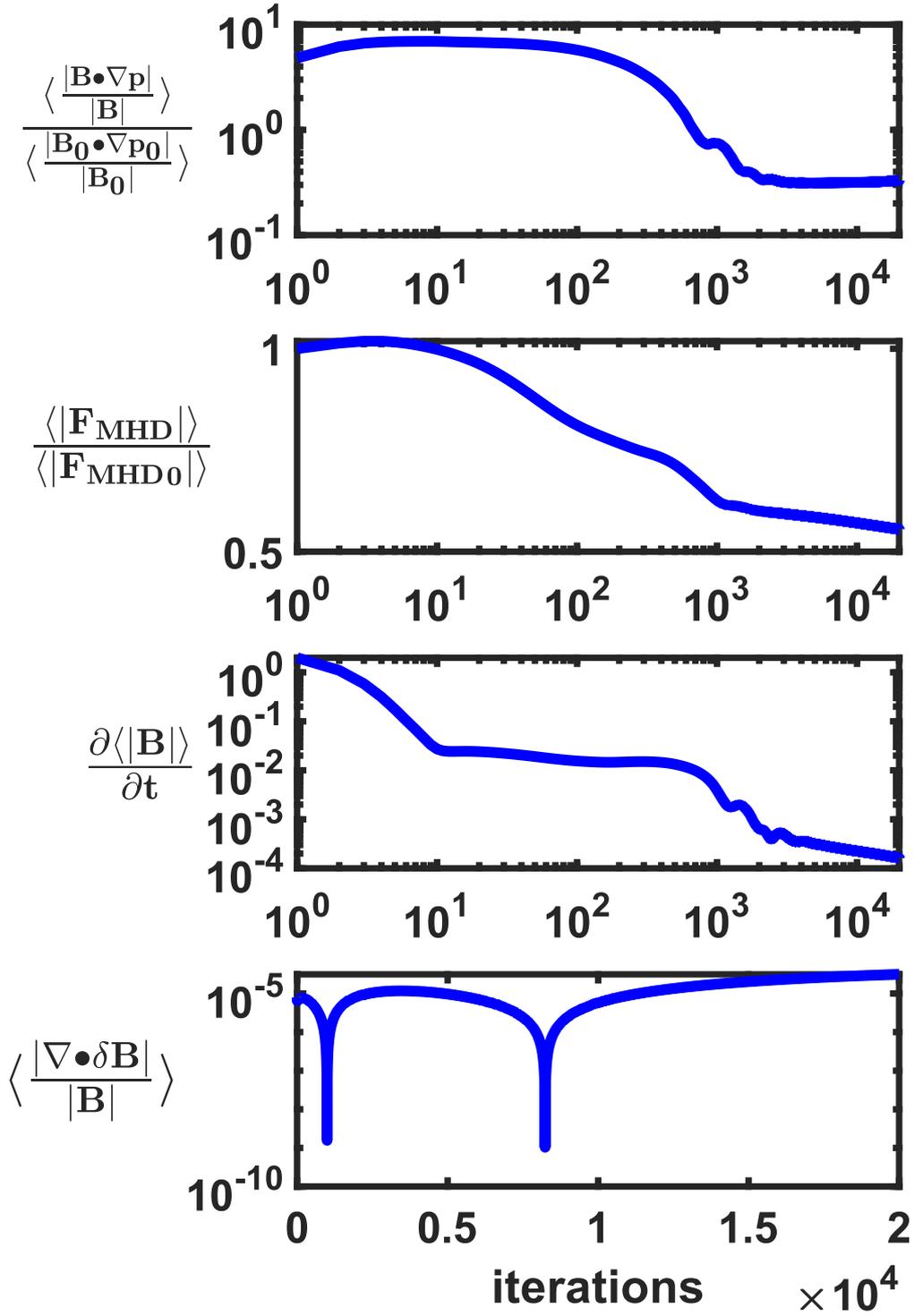

Fig. 8. Four convergence metrics versus iteration steps of the free-boundary CFQS equilibrium with $\eta = 1 \times 10^{-4}$. The calculation stops at the $20000^{th}$ iteration with $\langle|F_{MHD}|\rangle/\langle|F_{MHD_0}|\rangle \approx 0.55$ and $\langle \partial_t |B| \rangle \approx 1.62 \times 10^{-4}$.

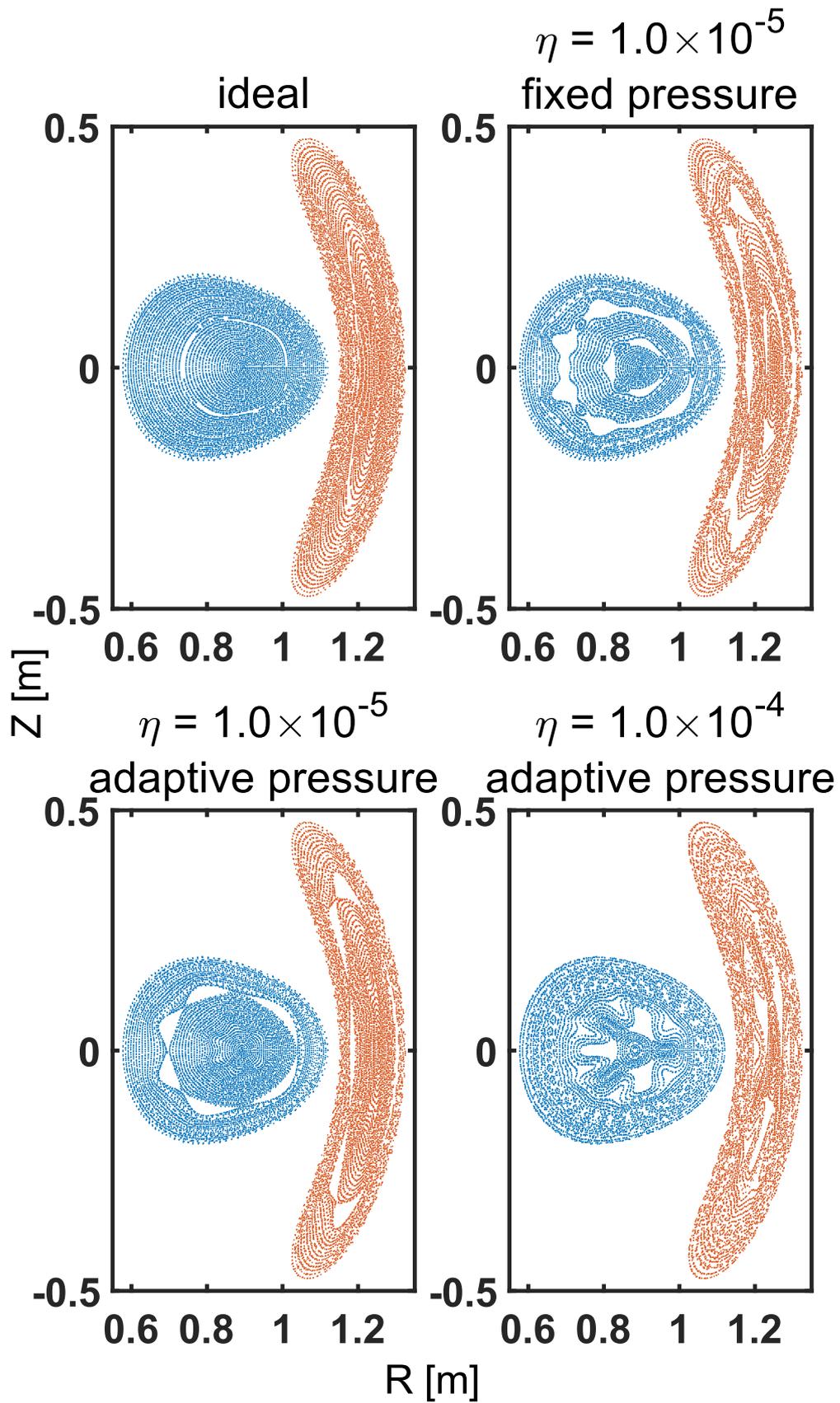

Fig. 9. Poincare plots at the cross sections $\varphi = 0$ (red) and $\pi/2$ (blue) of the initial ideal and three resistive CFQS fixed-boundary equilibria.

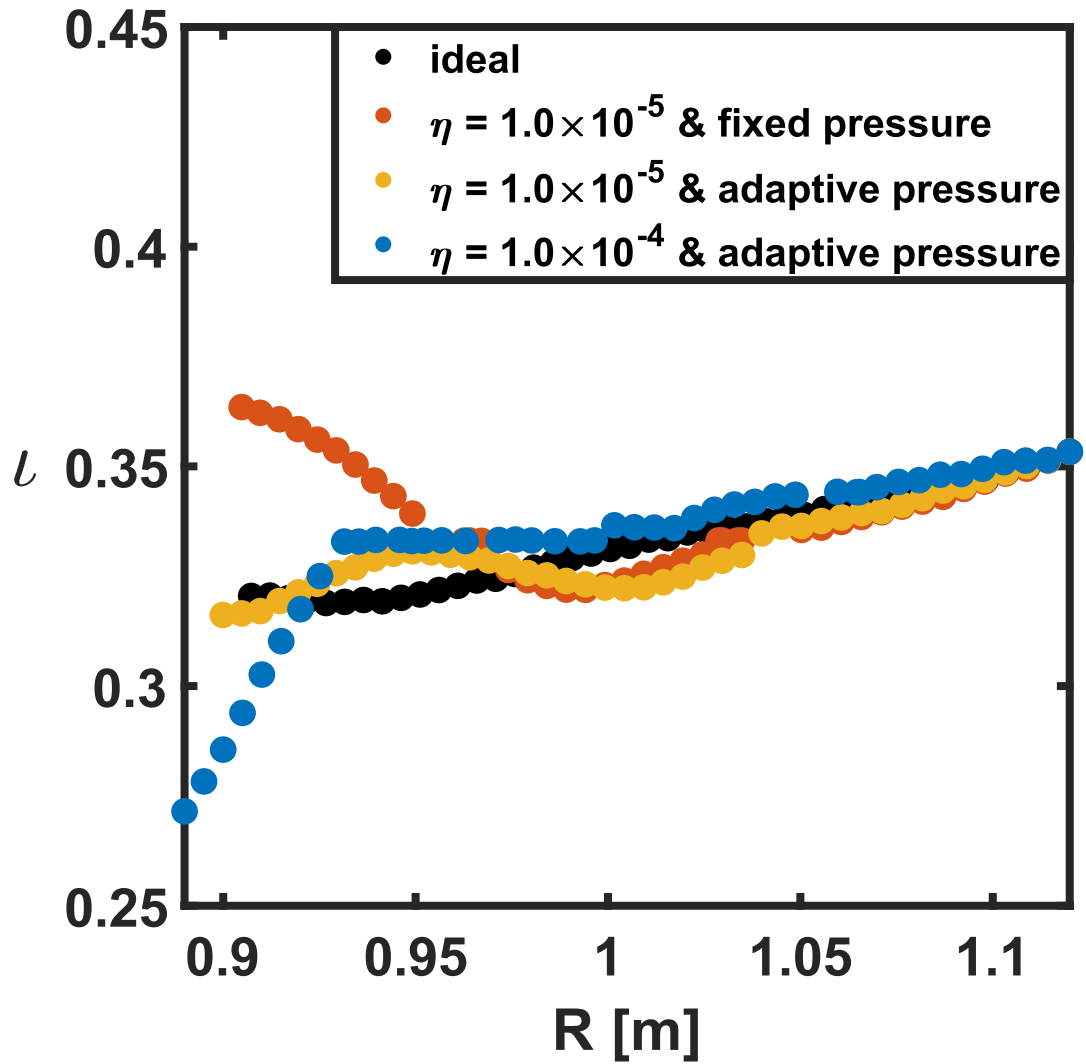

Fig. 10. Rotational transform profiles of the initial ideal and three resistive CFQS fixed-boundary equilibria.

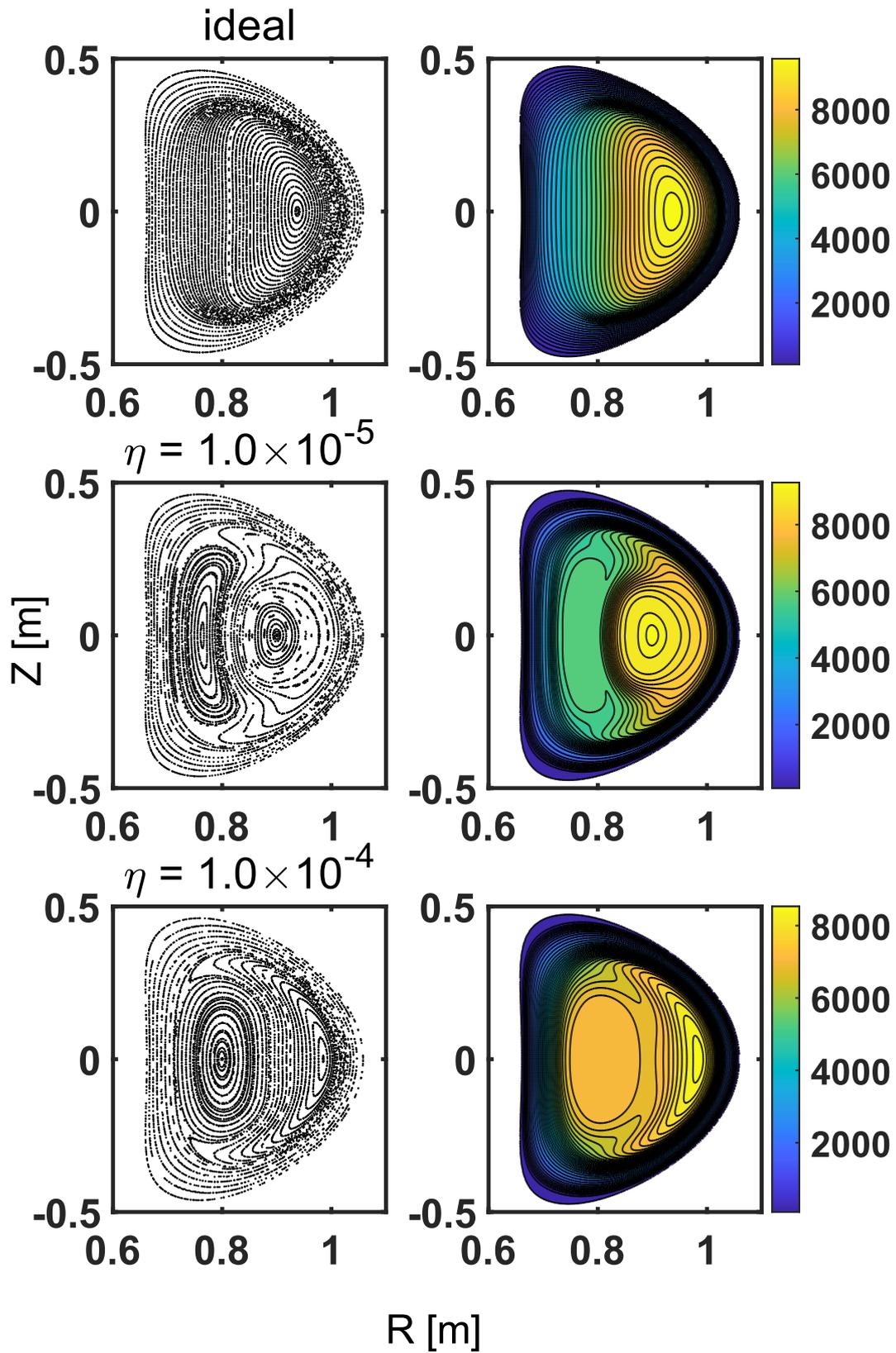

Fig. 11. Poincare plots (left column) and pressure isosurfaces (right column) at the cross section $\varphi = 0$ of the initial ideal and two resistive equilibria for TCV.

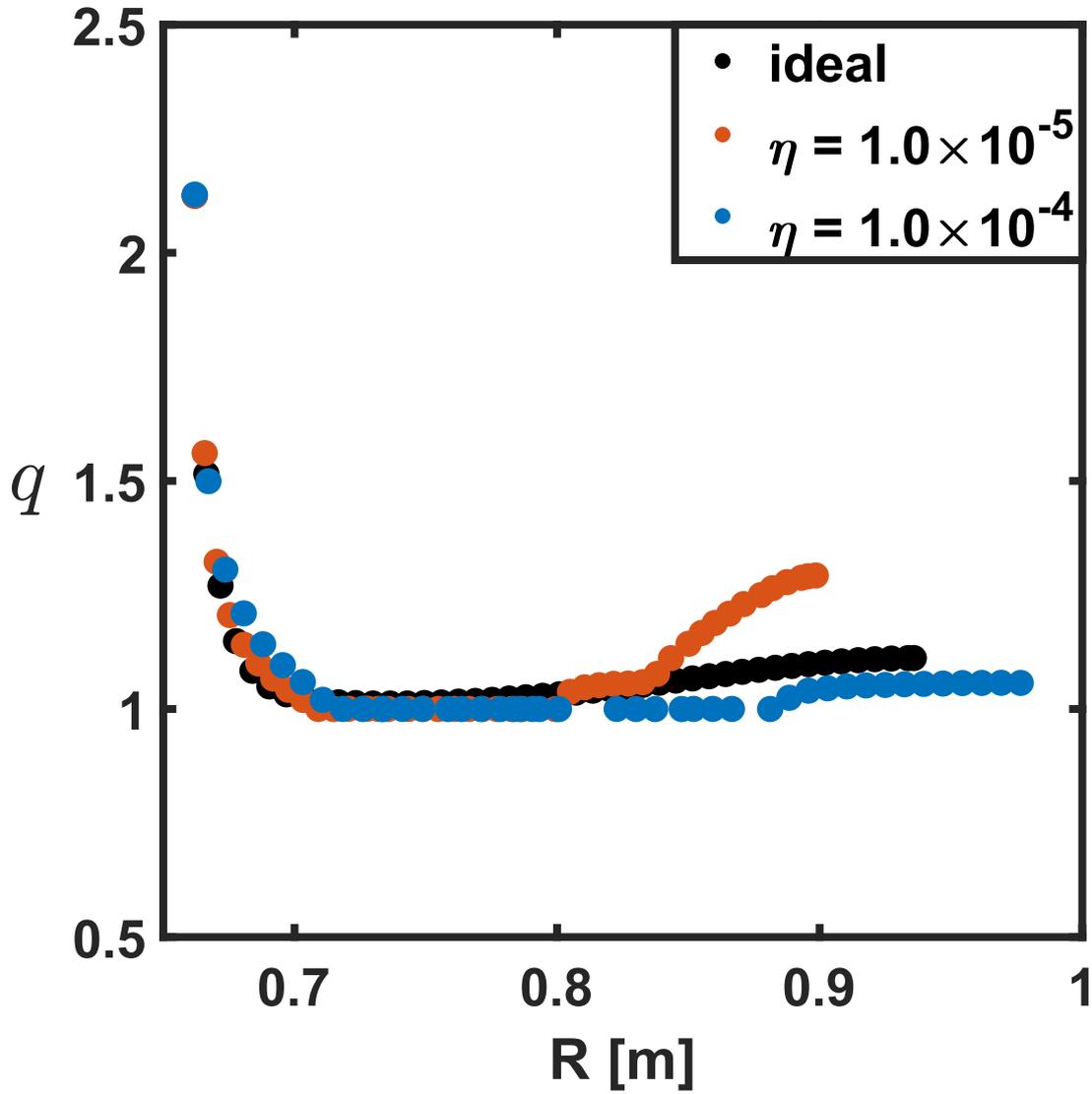

Fig. 12. Safety factor profiles illustrated at the cross section $\varphi = 0$ for two resistive TCV-like equilibria. Note that the left end of the abscissa corresponds to the minimum major radius of the last closed flux surface.

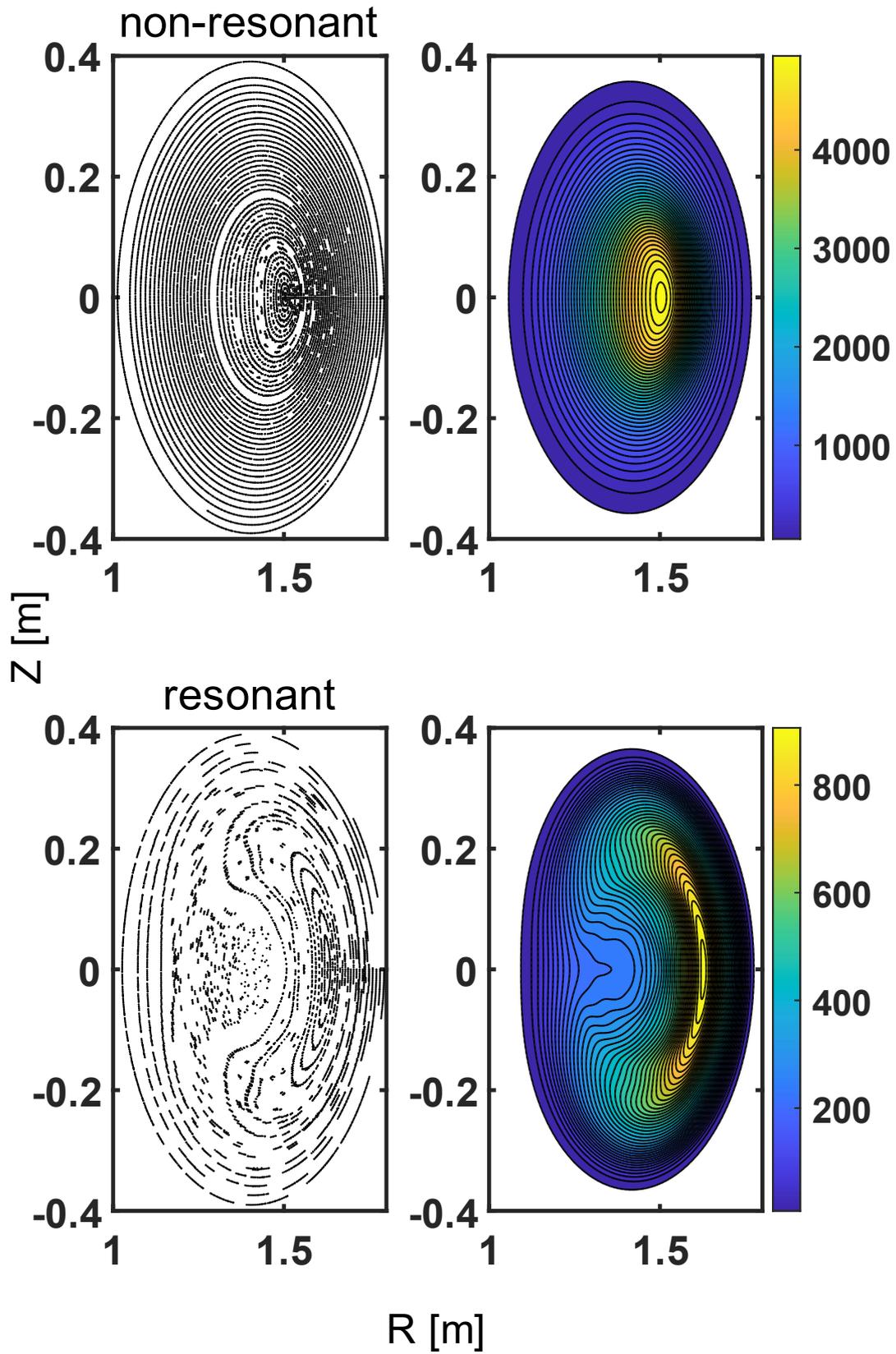

Fig. 13. Poincare plots (left column) and pressure isosurfaces (right column) at the cross section $\varphi = \pi/6$ of the KTX-QSH equilibria.

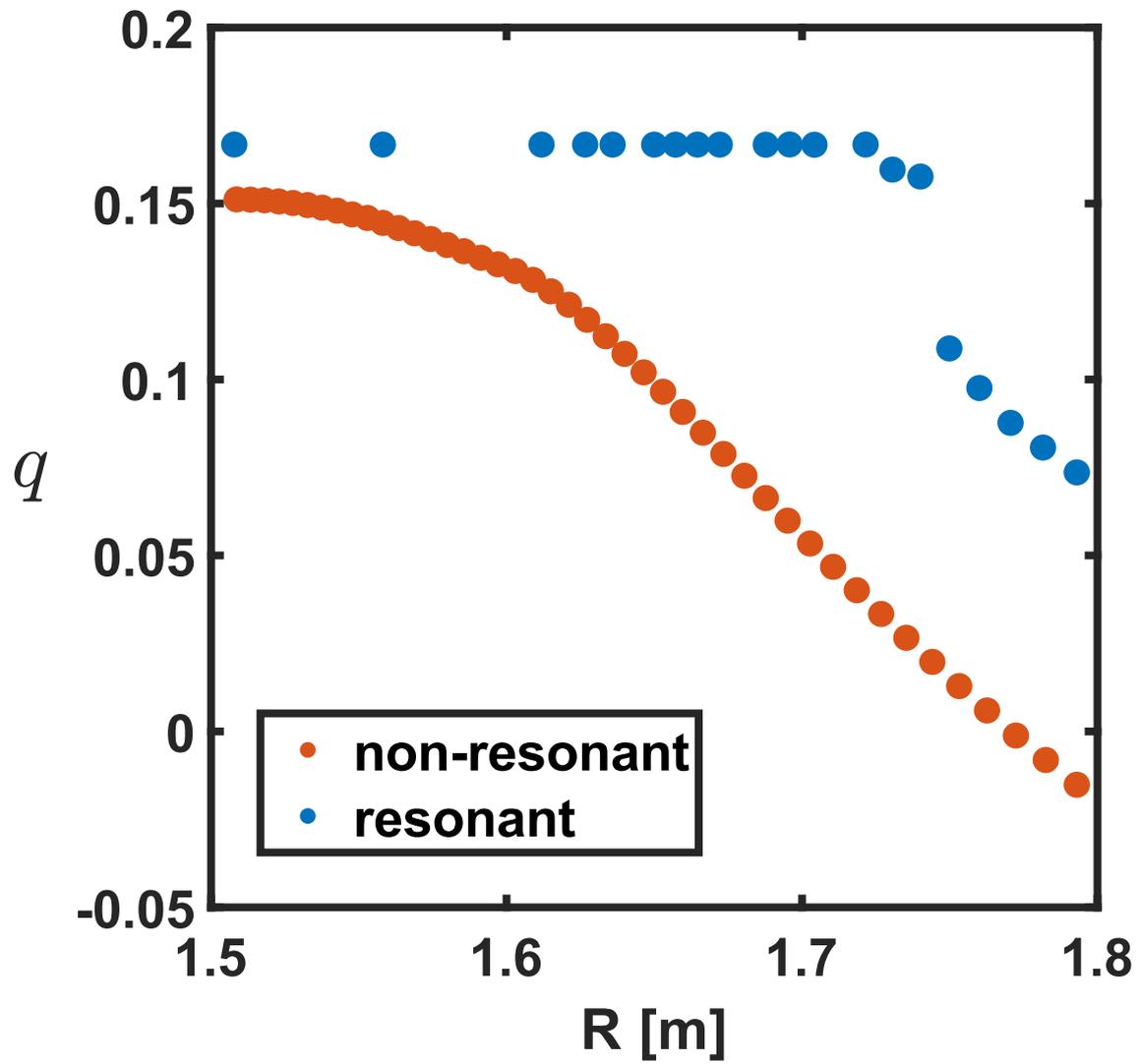

Fig. 14. Safety factor profiles of the corresponding resistive equilibria in Fig. 13